\documentclass[11pt,a4paper]{article}
\usepackage{epsfig}
\usepackage[T1]{fontenc}    
\usepackage{graphics}
\usepackage{graphicx}
\usepackage{pstricks,pst-coil,pst-fill,pst-plot}
\usepackage{cite,indentfirst}
\usepackage{amsmath}    
\usepackage{amssymb}    
\usepackage{amsfonts}   
\usepackage{verbatim}   
\usepackage{mathrsfs}   
\usepackage{dsfont}
\usepackage{vmargin}        

\setmarginsrb{3.4cm}{2.5cm}{3.4cm}{2.5cm}{1cm}{1cm}{1cm}{1.6cm}

\def\C|{{\mathbb C} \,}
\def\B|{{\mathbb B} \,}
\def\S|{{\mathbb S} \,}
\def\G|{{\mathbb G} \,}
\def\N|{{\mathbb N} \,}
\def\F|{{\mathbb F} \,}
\def\K|{{\mathbb K} \,}

\def\sg{\sigma}

\def\eps{\varepsilon}

\let\tend=\rightarrow

\newtheorem{prop}{Proposition}[section]

\newcommand\beq{\begin{equation}}
\newcommand\enq{\end{equation}}

\def\beqa{\begin{eqnarray}}
\def\eeqa{\end{eqnarray}}
\def\ba{\begin{array}}
\def\ea{\end{array}}

\def\a{\alpha}

\def\det{\operatorname{det}}
\def\eps{\epsilon}

\def\la{\lambda}
\def\sg{\sigma}
\def\pl{\prod\limits}
\def\lt({\left(}
\def\rt){\right)}

\numberwithin{equation}{section}

\newcommand{\f}[2]{{\ensuremath{%
    \mathchoice%
    {\dfrac{#1}{#2}}
    {\dfrac{#1}{#2}}
    {\frac{#1}{#2}}
    {\frac{#1}{#2}}
}}}
\newcommand{\tf}[2]{\ensuremath{#1/#2}}
\newcommand{\pa}[1]{\ensuremath{\left(#1\right)}}
\newcommand{\paa}[1]{\ensuremath{\left\{#1\right\}}}
\newcommand{\pac}[1]{\ensuremath{\left[#1\right]}}
\newcommand{\paf}[2]{\ensuremath{\left(\f{#1}{#2}\right)}}

\def\eps{\epsilon}

\def\la{\lambda}
\def\sg{\sigma}
\def\hx{\hat{\xi}}
\def\th{\theta}

\newcommand{\mc}[1]{\ensuremath{\mathcal{#1}}}

\newcommand{\ov}[1]{\ensuremath{\overline{#1}}}

\newcommand{\Int}[2]{\ensuremath{\int\limits_{#1}^{#2}}}

\newcommand{\sul}[2]{\ensuremath{\sum\limits_{#1}^{#2}}}

\newcommand{\R}{\ensuremath{\mathbb{R}}}

\newcommand{\Ctr}{\ensuremath{\mathscr{C}}}

\newcommand{\s}[1]{\ensuremath{\sinh\pa{#1}}}
\newcommand{\sd}[1]{\ensuremath{\mathfrak{s}\pa{#1}}}
\newcommand{\sdp}[2]{\ensuremath{\mathfrak{s}^{#1}\pa{#2}}}
\newcommand{\cd}[1]{\ensuremath{\mathfrak{c}\pa{#1}}}
\newcommand{\ch}[1]{\ensuremath{\cosh\pa{#1}}}

\newcommand{\ex}[1]{\ensuremath{\e{e}^{#1}}}

\renewcommand{\det}[2]{\ensuremath{\text{det}_{#1}\pac{#2}}}

\newcommand{\abs}[1]{\ensuremath{\mid #1 \mid}}

\newcommand{\moy}[1]{\ensuremath{\langle #1 \rangle}}

\newcommand{\Dp}[2]{\ensuremath{\f{\partial #1}{\partial #2}}}

\newcommand{\dd}{\text{d}}
\newcommand{\e}[1]{\ensuremath{\text{#1}}}

\newcommand{\intff}[2]{\ensuremath{\left [ \, #1 \,; #2 \, \right ] }}

\newcommand{\intof}[2]{\ensuremath{\left ] \, #1 \,; #2 \, \right ] }}
\newcommand{\intoo}[2]{\ensuremath{\left ] \, #1 \,; #2 \, \right [ }}

\newcommand{\intn}[2]{\ensuremath{[\![ \, #1 \,;\, #2 \,]\!]}}

\begin{document}


\author{K.~K.~Kozlowski\footnote{ Laboratoire de Physique, UMR 5672 du CNRS,  ENS Lyon,  France,
 karol.kozlowski@ens-lyon.fr}}

\begin{flushright}
LPENSL-TH-07/07\\
\end{flushright}
\par \vskip .1in \noindent

\vspace{24pt}

\begin{center}
\begin{LARGE}
{\bf   On the emptiness formation probability \\ of the open XXZ
spin-$\tf{1}{2}$ chain}
\end{LARGE}

\vspace{50pt}
\begin{large}
{\bf K.~K.~Kozlowski}\footnote[1]{ Laboratoire de Physique, ENS Lyon et CNRS,  France,
 karol.kozlowski@ens-lyon.fr}
\end{large}

\end{center}

\vspace{3cm}

\begin{abstract}
This paper is devoted to the study of the emptiness formation
probability $\tau\pa{m}$ of the open XXZ chain. We derive
 a closed form for $\tau\pa{m}$ at
$\Delta=\tf{1}{2}$  when the boundary field vanishes. Moreover we obtain its
leading asymptotics for an arbitrary boundary field at the free fermion point. Finally, we
compute the first term of the asymptotics of $\ln\pa{\tau\pa{m}}$ in the whole massless regime
$-1<\Delta<1$.
\end{abstract}

\newpage


\section{Introduction}

For already a few decades, spin chains trigger a lot of interest
as many relevant physical quantities can be computed.
For instance the spectrum (\!\cite{BetheSolutionToXXX}, \cite{Yang-YangXXZproofofBetheHypothesis})
of the spin-$\tf{1}{2}$ XXZ chain as well as its correlation functions
(\!\cite{JimKKKM92ElementaryBlocksXXZperiodicDelta>1}, \cite{KMTElementaryBlocksPeriodicXXZ})
 are known. The open
XXZ spin-\tf{1}{2} chain with diagonal boundary conditions:
\beq\label{HamXXZ}
  \mc{H}=\sum_{m=1}^{M-1}  \sigma^x_m \sigma^x_{m+1} +
  \sigma^y_m\sigma^y_{m+1} + \Delta\pa{\sigma^z_m\sigma^z_{m+1}-1}
+\sg^z_1 \, h_- + \sg^z_M \, h_+,
\enq

\noindent is another example of an integrable spin chain
 Its spectrum was first obtained by Alcaraz et \textit{al.}
\cite{AlcarazBAopenXXZ} in the framework of coordinate Bethe ansatz.
One year later, Sklyanin\cite{SklyaninABAopenmodels} developed an
appropriate scheme to apply the algebraic Bethe ansatz to models
with boundaries. In the mid nineties, the Kyoto group obtained an
integral representation for the elementary blocks of a half-infinite
massive ($\Delta>1$) XXZ spin-\tf{1}{2} chain with a diagonal
boundary field \cite{JimKKKM95XXZChainWithaBoundary}. Recently,
the Lyon group derived formulas for the elementary blocks
of the finite open chain XXZ with diagonal boundary fields and for
an arbitrary anisotropy parameter \cite{KozElemntaryblocksopenXXZ}. \\

The simplest possible correlation function of a spin-$\tf{1}{2}$
chain is the so-called emptiness formation probability (EFP)
$\tau\pa{m}$ first introduced in \cite{KorepinFirstIntroductionoftheEFP}. This quantity can be understood
as the probability of observing a ferromagnetic
string of length $m$ starting from the first site of the chain.
The EFP can be computed to the end in the case of a periodic XXZ spin-$\tf{1}{2}$ chain at $\Delta=\tf{1}{2}$.
Indeed this exact expression has been conjectured in \cite{StronRazuPeriodicEmptinessPiOver3}
and proven in \cite{KMNTPeriodicEmptinessPiOver3}. The free fermion point of the bulk model is also
quite particular as the EFP can be represented as a Toeplitz determinant of a smooth function on an
arc. Hence its asymptotics can be studied \cite{ShiroisihiTakaNishiEFPatDelta=0Periodic} by using the strong Szegö
limit theorem for circular arcs \cite{WidomSzegoLimitonCircularArcs}. One can also derive the
leading asymptotics of $\ln \tau\pa{m}$ of the bulk massless model using a
saddle point method \cite{KMNTPeriodicEmptinessAsymptAllDelta}. \\

The aim of this paper is to obtain corresponding results
for the open XXZ spin $\tf{1}{2}$ chain. The
symmetrized integral representation for the EFP derived in  \cite{KozResummationsOpenXXZ} will be the
starting point of our considerations. We first show that
$\tau\pa{m}$ can be computed to the end in the case of an open
spin-\tf{1}{2} chains subject to a zero boundary field at $\Delta=\tf{1}{2}$. After
setting some notations in Section \ref{Section de definitions }, we
will present this result in Section \ref{Section EFP at Delta=1/2}.
We then derive the leading asymptotics of the EFP at the free
fermion point in Section \ref{Section Leading asym of EFP Delta=0}.
This is done by exploiting the almost Hankel determinant structure
of the EFP at $\Delta=0$. We analyse the asymptotic behavior of such determinants in
Appendix \ref{appendice asymptotique determinant}. This analysis takes advantage of
the asymptotics of Toeplitz matrices with Fischer-Hartwig type singularities
\cite{EhrhardtAsymptoticBehaviorOfFischerHartwigToeplitzGeneralCase} as well as
of the uniform asymptotics of orthogonal polynomials of modified Jacobi type
\cite{KuilajaarsMVVUniformAsymptoticsForModifiedJacobiOrthogonalPolynomials}.
In  the last section  we use the saddle point approximation  borrowed from the bulk
\cite{KMNTPeriodicEmptinessAsymptAllDelta} to derive the leading asymptotics
of $\ln \tau \pa{m}$ in the massless regime.

\section{The Emptiness formation probability.}
\label{Section de definitions }
 The authors of \cite{KozResummationsOpenXXZ} showed that, in the case of
 a half-infinite chain, the emptiness formation probability admits a multiple integral
representation whose  integrand is a symmetric function of the integration variables.
 This way of representing $\tau\pa{m}$ enables
the separation of variables in the $m$-fold integral defining
$\tau\pa{m}$ at $\Delta=\tf{1}{2}$. It also allows to express
$\tau\pa{m}$ as an almost Hankel determinant at $\Delta=0$.
Finally, it is suited for the derivation the leading asymptotics of
$\ln\pac{\tau\pa{m}}$ in the massless regime.

We recall the standard parameterizations of the massless regime
\beq
\Delta=\cos \pa{\zeta}\, \;, \;  \zeta\in \intoo{0}{\pi}\,  \quad \e{ and } \quad
%
        h_-=-\sin\zeta \cot{\hx_-}  \, \; \; , \;\;
         \hx_-\in \intof{-\tf{\pi}{2}}{\tf{\pi}{2}}
\nonumber
\enq

\noindent as well as the symmetrized integral representation for the
emptiness formation probability \cite{KozResummationsOpenXXZ}:
\beqa
\hspace{-2cm}\tau\pa{m}&=&\moy{E_{22}^{(1)}\dots E_{22}^{(m)}} \nonumber\\
&=&\f{1}{2^m\, m! } \Int{\mc{C}\pa{h}}{} \dd^m \la
 \f{\text{det}_m\pac{\Phi\pa{\la_j,\xi_k}} \, \text{det}_m\pac{\tf{1}{h\pa{\la_j,\xi_k}}}}{\pl_{i<j}
\sdp{2}{\xi_i,\xi_j}}  \nonumber \\
&&  \hspace{1cm} \times \f{\pl_{i,j}^{m}
h\pa{\la_i,\xi_j}}{\pl_{j>k}^{m}\sd{\la_{jk},i\zeta}\sd{\ov{\la}_{jk},i\zeta}}
\, \pl_{k=1}^{m} \f{\s{2\la_k}
\sd{\xi_k,i\hx_-}}{\sd{\la_k,i\hx_-+i\tf{\zeta}{2}}}
\label{emptiness for general zeta}
\eeqa

The contours of this multiple integral depend on the boundary magnetic field
$h_-$:
\beq
\mc{C}\pa{h}= \R \underset{\pm}{\cup}
\Gamma_{\pm}\paa{\mp i\pa{\hx_- + \tf{\zeta}{2}}} \;
\e{whenever}
 \; -\tf{\zeta}{2}<\hx_-<0 \; \e{and} \; \mc{C}\pa{h}= \R \; \e{otherwise}.
\nonumber \enq

\noindent Here
$\Gamma_{\pm}$ stands for a small circle skimmed through in the
anticlockwise/clockwise direction. Furthermore, we agree that
\beq
    \ba{ccc}
         \sd{x,y}=\s{x+y} \s{x-y} &\qquad &\cd{x,y}=
            \ch{x+y}\ch{x-y}    \vspace{4mm}\\
            \la_{ij}=\la_i-\la_j &\qquad &  \ov{\la}_{ij}=\la_i+\la_j
    \ea
\enq

\noindent and introduce the functions:
\beqa
\Phi\pa{\la,\xi}&=& \f{\rho\pa{\la-\xi}-\rho\pa{\la+\xi}}{\s{2\xi}}
= \f{\s{\tf{\pi \la}{\zeta}} \s{\tf{\pi \xi}{\zeta}}}{\zeta
\s{2\xi}\mathfrak{c}\pac{
\f{\pi}{\zeta}\pa{\la,\xi}}}  \\
h\pa{\la, \xi} &=& \sd{\la+\xi,i\tf{\zeta}{2}}
\sd{\la-\xi,i\tf{\zeta}{2}}
\eeqa

\noindent Lastly, we remind that $\rho\pa{\la}= \f{1}{\zeta
\ch{\tf{\pi\la}{\zeta}}}$ solves the Lieb equation:
\beq
\f{\sin\zeta}{ \pi \sd{\la,i\tf{\zeta}{2}}}= \rho\pa{\la} +
\Int{\R}{}\dd \mu \,  K\pa{\la-\mu}\, \rho\pa{\mu} \,\,\, ; \qquad
 K\pa{\la}= \f{\sin 2\zeta}{2\pi\, \sd{\la,i\zeta}}
\label{equation de Lieb}
 \enq
 
\section{$\zeta=\tf{\pi}{3}$, an exact result.}
\label{Section EFP at Delta=1/2}

 When $\zeta=\tf{\pi}{3}$, the integrand of eq.\eqref{emptiness for general zeta} can be further simplified due to
the duplication formula:
\beq
\s{3x}=4 \s{x}\s{x+i\tf{\pi}{3}}\s{x-i\tf{\pi}{3}} \,\, .
\enq

\noindent Indeed, in this special point one recasts $\tau\pa{m}$ as
\beqa
\hspace{-2cm} \tau\pa{m}&=&\f{\pa{\tf{3}{4}}^{m\pa{m+1}}}{\pl_{i<j}\mathfrak{s}^2\pa{\xi_i,\xi_j}}
\Int{\mc{C}\pa{h}}{} \!\!\f{\dd^m \la}{m! \pa{4\pi}^m} \pl_{k=1}^{m}
\f{\s{3\la_k}\s{2\la_k}
\sd{\xi_k,i\hx_-}}{\sd{\la_k,i\hx_-+i\tf{\pi}{6}}}  \nonumber \\
&& \hspace{1cm} \times \det{m}{\f{1}{\cd{\la_j,\xi_k}}} \det{m}{\f{1}
{h\pa{\la_j,\xi_k}}}
\eeqa

\noindent  A partial
homogeneous limit has already been performed in the latter formula.

The symmetry of the integrand enables us to replace the Cauchy
determinant \newline $\det{m}{\tf{1}{\cd{\la_j,\xi_k}}}$ by  $m!$ times the
product of its diagonal elements. This yields the sought separation
of variables. In the case of a zero boundary field
( {\it ie.} $\xi_-=i\tf{\pi}{2}$), the evaluation of the integrals leads to

\beq
\tau\pa{m}= \paf{3}{4}^{m\pa{m+1}}
\f{\pa{-2}^m}{\pl_{k=1}^{m}\s{2\xi_k}\s{\xi_k}}
\f{\det{m}{g\pa{\xi_j,\xi_k}}}{\pl_{j<k}\mathfrak{s}^2\pa{\xi_j,\xi_k}}\;\; ,
  \label{tauinhomogeneous}
\enq

\noindent where
\beq
g\pa{x,y}=\f{\sinh\tf{\pa{x+y}}{2}}{\sinh3\tf{\pa{x+y}}{2}}
-\f{\sinh\tf{\pa{x-y}}{2}}{\sinh3\tf{\pa{x-y}}{2}}
\enq

It happens that it is possible to evaluate the homogeneous limit of \eqref{tauinhomogeneous}.
Indeed, we have the following equality of limits:
\beq
\lim_{\xi_k\rightarrow 0}
\f{\det{m}{g\pa{\xi_j,\xi_k}}}{\pl_{k=1}^{m}\xi^2_k \pl_{i<j}^{m}
\pa{\xi^2_j-\xi^2_k}^2 } =
\lim_{\substack{x_i\tend 0 \\ y_i \tend 0}}
\f{\det{m}{g\pa{x_j,y_k}}}{\pl_{k=1}^{m}x_k\,y_k \pl_{i<j}^{m}
\pa{x^2_j-x^2_k}\pa{y^2_j-y^2_k} } \,\,.
\label{egalite limites}
 \enq

\noindent We can send the $x$'s and the
$y$'s to zero in an arbitrary way when computing the limit in the RHS of \eqref{egalite limites}.
 In particular, we can choose $x_i=\a \, i$ and $y_j=\a \, \pa{m+j}$, and sent $\a$  to $0$. Such a
homogeneous limit is the $\a \tend 0$ limit
 of $\det{m}{U}$, where $U \in \mc{M}_m\pa{\mathbb{C}}$   is given by
\beq
U_{i,j}= \f{\sinh \a \tf{\pa{i+j+m}}{2}}{\sinh \beta
\tf{\pa{i+j+m}}{2}} -\f{\sinh \a \tf{\pa{j-i+m}}{2}}{\sinh \beta
\tf{\pa{j-i+m}}{2}} \,\,  \qquad \beta = 3 \a.
\enq

\noindent Actually Kuperberg \cite{KuperbergSymclassOfA.M.S.} evaluated $\det{m}{U}\;
$ for all $ \a$  and $\beta$:
\beqa
\det{m}{U}=\f{\pl_{i<j}^{2m}2 \sinh \beta\tf{\pa{j-i}}{2} \pl_{\substack{i,j\\
2 \mid j}}^{2m+1} 2 \sinh\tf{\pa{\a+\beta\pa{j-i}}}{2}}
{\pl_{i,j}^m 4 \sinh\tf{\beta\pa{m+j-i}}{2}
\,\,\, \sinh\tf{\beta\pa{m+j+i}}{2}}
\label{Kuperbergdeterminant} \;\; .
\eeqa

\noindent Hence:
\beqa
\lim_{\xi_k\rightarrow 0}
\f{\det{m}{g\pa{\xi_j,\xi_k}}}{\pl_{k=1}^{m}\xi^2_k \pl_{i<j}^{m}
\pa{\xi^2_j-\xi^2_k}^2 } =
\f{1}{\pl_{\substack{  i<j \\ 1\leq}}^{m}
\pa{i^2-j^2}\pl_{\substack{ \leq i<j \\ m+1}}^{2m}\pa{i^2-j^2}}
\lim_{\a\rightarrow 0} \a^{-2m^2}\f{\det{m}{U}}{2m!}
\eeqa

\noindent This allows to obtain a closed formula for the emptiness formation
probability in the homogeneous limit:
\beq
\tau\pa{m}= \f{ \pa{\tf{3}{4}}^{m^2}}{\pa{-2}^m\pa{2m}!}
\f{    \pl_{i<j}^{2m} \pa{j-i} \pl_{\substack{i,j \\
2\mid j}}^{2m+1} 1+3\pa{j-i}}
{\pl_{i,j}^{m} \pa{m+j-i}\pa{m+j+i} \pl_{i<j}^{m}\pa{j^2-i^2}\pl_{\substack{\leq i<j\\  m+1}}^{2m}\pa{j^2-i^2}} \, .
\enq

\noindent Or in terms of factorials:
\beq
\tau\pa{m}=3 \f{2^{m+1}}{4^{m^2}} \pl_{j=1}^{m} \pa{3j-1}
\pl_{k=1}^{m-1} \f{\pa{6k+3}!}{\pa{2k+1}!}
\paa{\f{1}{\pa{2m}!} \pl_{k=1}^{2m} \f{k!}{\pa{2m+k}!}}^{\f{1}{2}} \; .
\label{tau(m)factorial}
\enq

It is now a matter of standard asymptotic analysis to extract the large $m$ behavior of
\eqref{tau(m)factorial}:

\beq
\tau\pa{m}= \paf{3}{4}^{3m^2} \paf{3 \sqrt{3}}{4}^m m^{\f{1}{72}} \, C \,\, \pa{1+\text{o}\pa{1}}\;\; ,
\enq

\noindent where the constant $C$ is expressed in terms of
the Euler Gamma function $\Gamma\pa{z}$, the
Euler constant $\gamma$ and of the Riemann Zeta function$\zeta\pa{z}$.

\beqa
C&=& \f{2^{\f{1}{4}}\sqrt{2}}{\Gamma\pa{\tf{2}{3}}} \exp\pa{\f{9}{2} \zeta'\pa{-1}+\f{5}{36}\pa{1+\gamma}}
 \\
&&\exp\paa{2\Int{0}{+\infty} \dd t  \f{\e{e}^{-t}}{1-\e{e}^{-t}}
\pa{\f{2-\cosh\tf{t}{3}-\cosh\tf{t}{6}}{t^2}
+\f{2\sinh\tf{t}{3}+\sinh\tf{t}{6}}{6t}
-\f{5}{72}}}   \;\; . \nonumber
\eeqa

We stress that the qualitative behavior of $\tau\pa{m}$ differs from the bulk one by the
presence of an exponential factor $\ex{c m}$ in the asymptotics . This seems to be a
 general feature of the boundary model as a similar behavior appears at the free fermion point.
Moreover the gaussian decay is twice faster than in the bulk.

\section{The leading asymptotics of $\tau\pa{m}$ at $\Delta=0$}
\label{Section Leading asym of EFP Delta=0}

The integral representation for the EFP also admits a separation of
variables at $\Delta=0$ (\textit{ie} $\zeta=\tf{\pi}{2}$). More
precisely  specifying eq.\eqref{emptiness for general zeta} to $\zeta=\tf{\pi}{2}$
we get

\beq
\tau\pa{m}= \f{2^{m\pa{m+1}^2}
\sin^{2m}\!\pa{\hx_-}}{\pa{2\pi}^m} \text{det}_m\pac{T}
\enq

\noindent The entries of the matrix $T$ are defined by an
integral:

\beq
T_{jk}=\Int{\mc{C}\pa{h_-}}{}
\f{\sinh^2\pa{2\la}}{\cosh^{k+l}\pa{2\la}\pac{\cosh \pa{2\la} +
\sin\pa{2\hx_-}}} \; \dd \la
\enq

We perform the change variables $z=\tf{2}{\cosh\pa{2\la}}-1$ in the above integral. Then $\tau\pa{m}$ reads

\beq
\tau\pa{m}=
\f{\text{det}_m\pac{R}}{\pa{-2\pi h_-}^m} \hspace{1cm } \e{where} \hspace{1cm}
R_{jk} =   \ex{i\f{\pi}{2}} \lim_{\eps \tend 0^+}\Int{\mathscr{C}_{\eps}}{}   \f{z^{j+k-2}}{z-a}
                    \, \omega\pa{z} \, \dd z \;\;\; ,
\enq

and

\beqa
&&\hspace{1cm} \mathscr{C}_{\eps}=  \left\{ \ba{cc} \intoo{-1-i\eps}{1-i\eps} \bigcup \Gamma_{+}\pa{a}  & h_- \in \intoo{1}{+\infty}\\
                                    \intoo{-1-i\eps}{1-i\eps}   & h_- \in \intoo{-\infty}{1}
                            \ea \right. \\
&& \omega\pa{z}=  \sqrt{\pa{z-1}\pa{3+z}} \qquad \e{and} \qquad a=h_-+h^{-1}_- - 1\equiv \f{\chi+\chi^{-1}}{2}\nonumber
\eeqa

In order to apply the results of Appendix \ref{appendice asymptotique determinant} we 
recall the Wiener-Hopf factorization of $\sqrt{3+\cos \th}$
\beq
\sqrt{3+\cos \th} = \sqrt{\f{u}{2}}\pa{1+u^{-1}\ex{i\th}}^{\f{1}{2}}\pa{1+u^{-1}\ex{-i\th}}^{\f{1}{2}} \; .
\enq

Specializing eq. \eqref{Asymptotique de Det R} to the determinant representation for the EFP, we get
\beq
\tau\pa{m} \sim C \f{u^{\f{m}{2}}}{2^{m^2}}\paf{2}{m}^{\f{1}{8}}
    \left\{ \ba{c c} \pa{h_-\chi}^{-m} \sqrt{\f{1+\chi^{-1}u^{-1}}{1+\chi^{-1}}}, \quad & h_-<1\vspace{4mm}\\
        \paf{\chi}{h_-}^m \sqrt{\f{\chi^{-1}+u^{-1}}{1+\chi^{-1}} }, & h_->1  \ea \right. \;\; .
\enq

The constant C is expressed in terms of $u$ and of the Barnes G-function
\beq
C=\paf{1+u^{-1}}{1-u^{-1}}^{\f{1}{8}}
\f{\pi^{\f{1}{4}}G\pa{\tf{1}{2}}}{\sqrt{1+u^{-1}}}
\enq

One can see explicitly from the large $m$ asymptotics of $\tau\pa{m}$ the asymmetry induced by the
boundary field. This asymmetry appears for $\abs{h_-}>1$. One has

\beq
\f{\tau_{h_-}\pa{m}}{\tau_{-h_-}\pa{m}} = \sqrt{\f{\pa{u+\chi^{-1}_+}\pa{1+\chi^{-1}_-}}
                {\pa{u\chi^{-1}_-+1}\pa{1+\chi_+^{-1}}}}  \pa{-\chi_+\chi_-}^{m}  \qquad h_- >1
\enq

\noindent where
\beq
\f{\chi_{\pm}+\chi_{\pm}^{-1}}{2}=\pm\pa{h_-+h_-^{-1}}-1 \; .
\enq

Finally, one infers from the asymptotics that $\tau\pa{m}\underset{h_- \tend -\infty}{\longrightarrow}0$
for large $m$, as it should be. Indeed, in such a limit, the first spin is necessarily oriented upwards.
Conversely, when $h_-\tend +\infty$, $\tau\pa{m}$ is not vanishing, also as expected.

\section{The $m\tend +\infty$ limit  of $\tf{\ln\pa{\tau\pa{m}}}{m^2}$}
\label{Section Leading of tau of m all Delta}

Using the saddle point approximation we obtain the leading asymptotics of
$\ln\pa{\tau\pa{m}}$. We find that the leading terms is gaussian and independent of
the value of the boundary field $h_-$:

\beq
\ln\pa{\tau\pa{m}} = K m^2 +\e{o} \pa{m^2}
\enq

\noindent The constant K is twice bigger than the one appearing in the leading
asymptotics  of $\tau\pa{m}$ of the periodic chain
\cite{KMNTPeriodicEmptinessAsymptAllDelta}.

\noindent In order to implement the asymptotic analysis of $\ln \tau\pa{m}$ we
first recast $\tau\pa{m}$ as
\beq
\tau\pa{m}= \f{\abs{\s{\xi_-}}^{2m}}{\pa{2\zeta}^{2m}}
\paf{\pi}{\sin\pa{\zeta}}^m \paf{\pi}{\zeta}^{2m^2} \mc{I}_m
\enq

\noindent $\mc{I}_m$ is an $m$-fold integral 
\beqa
\mc{I}_m=\f{1}{2^m \, m!}\Int{\mc{C}\pa{h_-}}{} \dd^m \la
\,\, \e{e}^{m^2 \, \mathscr{S}\pa{\paa{\la}_1^m}}
\det{m}{M_{jk}}
\eeqa
\beqa
\mathscr{S}\pa{\paa{\la}_1^m}&=& \f{1}{m^2} \sul{k=1}{m}
\ln\paf{\sinh^2\tf{\pi \la_k}{\zeta}}
{\s{\la_k,\xi_-+i\tf{\zeta}{2}}}
+\f{1}{m^2} \sul{k<l}{} \ln\paf{\sdp{2}{\tf{\pi
\pa{\la_k,\la_l}}{\zeta}}}
{\sd{\la_{kl},i\zeta}\sd{\ov{\la}_{kl},i\zeta}} \nonumber \\
&& \qquad +\f{2}{m} \sul{k=1}{m} \ln
\paf{\sd{\la_k,i\tf{\zeta}{2}}}{\cosh^2 \tf{\pi \la_k}{\zeta}} \;\; ,
\eeqa

\noindent and the matrix $M$ is obtained by inverting the equation
\beq
\sul{b=1}{m} M_{ib} \Phi\pa{\la_b,\xi_k} = \f{\s{2\la_j} \sin
\zeta}{ \pi h\pa{\la_j,\xi_k}} \label{matriceM} \;\; .
\enq

\noindent We now reexpress $\mc{I}_m$ in a form more suited for
an asymptotic analysis:
\beqa
\mc{I}_m &=& \hspace{-5mm}\Int{0<\la_1<\dots<\la_m}{} \hspace{-7mm}
\dd^m \!\la
  \; \e{e}^{m^2 \,\mathscr{S}\pa{\paa{\la}_1^m}}
\det{m}{M_{jk}} \,\,\,
+ \delta_{\hx_-} \f{\sinh^2{\tf{\pi \la_m}{\zeta}}}{\s{2\la_m}}
\hspace{-5mm}\Int{\substack{0<\la_1<\dots<\la_{m-1}\\
-i\la_m=\tf{\zeta}{2}+\hx_-}}{} \hspace{-7mm} \dd^{m-1}\! \la
 \nonumber \\
&& %
 \,\, \times \e{e}^{m^2 \,\mathscr{S}\pa{\paa{\la}_1^{m-1}}}
 \e{det}_{m}
\underbrace{\pac{M_{jk} \f{
\sd{\la_m,i\tf{\zeta}{2}}\sdp{2}{\tf{\pi\pa{\la_m,\la_j}}{\zeta}}}
{\sd{\la_{jm},i\zeta} \sd{\ov{\la}_{jm},i\zeta} \cosh^4\pa{\tf{\pi
\la_m}{\zeta}}}}}_{\widetilde{M_{jk}}}
\label{prise en compte de la racine complexe}
\eeqa

Here
\beq
\delta_{\hx_-}= \left\{ \ba{c c} 
                        1 & \hx_- \in \intoo{-\f{\zeta}{2}}{0} \\
                        0 & \e{otherwise}  \ea \right .  \quad .
\enq

Note that we have used the symmetry in the $\la$'s as well as the
$\la \rightarrow -\la$ invariance of the integrand in order to
replace the integration over $\R^m$ by $2^m m!$ times the
integration over the ordered domain $\paa{0<\la_1<\dots < \la_m}$.
We also used the determinant structure and the symmetry of the
integrand to evaluate the contribution of poles, if it exists. \\

The leading contributions to the above integrals will be equal to
the
 integrand evaluated at the solutions of the saddle point equations.
 We assume that, in the $m\tend +\infty$ limit,
these solutions  densify  on $\R^+$ with a density $\sg\pa{\la}$. Then sums can be replaced by integrals
according to the prescription

\beq
\f{1}{m}\sul{i=1}{m} f\pa{\la_i} \underset{m \tend + \infty}{\longrightarrow}
\Int{0}{+\infty} \dd \la  \; \sg\pa{\la} \, f\pa{\la}
\enq

It is worth noticing that in this limit, $M_{jp}$ can be approximated by

\beq
M_{jp}= \delta_{jp} + \f{K\pa{\la_{jp}}-K\pa{\ov{\la}_{jp}}}{ m
\sg\pa{\la_j}}  \;\; .
\enq

\noindent This can be seen by replacing the sums by integrals in
\eqref{matriceM} and using the integral equation \eqref{equation de
Lieb}. Then, Hadamard's inequality

\beq
\abs{\det{m}{a_{jk}}} \leq
\pa{\underset{j,k}{\max}\pa{\abs{a_{jk}}}}^m m^{\f{m}{2}}
\enq

\noindent ensures that

\beq
\lim_{m\tend +\infty} \f{\ln \det{m}{M_{j,k}} }{m^2} = \lim_{m\tend
+\infty} \f{\ln \det{m}{\widetilde{M}_{j,k}} }{m^2}= 0 \,\, .
\enq

\noindent As a consequence, the functions $\det{m}{M}$
and $\det{m}{\widetilde{M}}$ cannot contribute to the leading asymptotics of $\ln \pa{\tau\pa{m}}$.

 The density of saddle point roots $\sg\pa{\la}$ satisfies a singular integral equation. This equation
is obtained by taking the large $m$ limit in the saddle point equations $\Dp{\mc{S}\pa{\paa{\la}}}{\la_j}=0$. 
This singular integral equation reads:
\beqa
&&2\pa{\f{2\pi}{\zeta} \tanh \f{\pi \la}{\zeta} -\coth\pa{\la+i\tf{\zeta}{2}}-\coth\pa{\la-i\tf{\zeta}{2}}}= \\
\label{eqnintegraledensityracines}
&& \qquad V.P. \Int{\R^+}{}  \dd\mu \sg\pa{\mu}  \sul{\eps=\pm}{}
 \f{2\pi}{\zeta} \tanh\pi \f{\la-\eps \mu}{\zeta}  - \coth\pa{\la-\eps\mu+i\zeta} \coth\pa{\la-\eps\mu-i\zeta}
\nonumber
\eeqa

It is natural to extend the density into an even function on $\R$.
This recasts the integral equation into a form very close to the
integral equation appearing in the bulk model\cite{KMNTPeriodicEmptinessAsymptAllDelta}:

\beqa
&&\f{2\pi}{\zeta} \tanh \f{\pi \la}{\zeta} -\coth\pa{\la+i\tf{\zeta}{2}}-\coth\pa{\la-i\tf{\zeta}{2}}= \\
\label{eqnintegraledensityracinesII}
&& \qquad V.P. \Int{\R}{}  \dd\mu \sg\pa{\mu}  \paa{
 \f{2\pi}{\zeta} \tanh\pi \f{\la- \mu}{\zeta}  - \coth\pa{\la-\mu+i\zeta} \coth\pa{\la-\mu-i\zeta}} \,\, .
\nonumber
\eeqa

Eq.\eqref{EqnintegraleSingulieredensiterecines} is solved by the Fourier transform. We find

\beq
\hat{\sg}\pa{k}= \f{2\, \cosh \tf{k \zeta}{2}}{\cosh k \zeta}
\qquad \e{hence} \qquad
\sg\pa{\la}= \f{\sqrt{2}\cosh\tf{\pi
\la}{2\zeta}}{\zeta\cosh\tf{\pi \la}{\zeta}}
\label{EqnintegraleSingulieredensiterecines}\enq

It is then immediate to evaluate  the $m\tend +\infty$ limit of
$\mathscr{S}\pa{\paa{\la}_1^{m}}$ at the solutions of the saddle
point equations:

\beq
\lim_{m \tend +\infty}\mathscr{S}\pa{\paa{\la}_1^{m}}=
\mathscr{S}^{(\infty)}=
\Int{\R-i0}{} \f{\dd k}{k}
\f{\cosh^2\pa{\tf{k \zeta}{2}} \,\,\sinh\pac{ k\tf{\pa{\pi-\zeta}}{2}} }
{\cosh \pa{k \zeta} \, \, \sinh\pa{\tf{k\zeta}{2}} \sinh\pa{ \tf{\pa{k\pi}}{2} }   } \,\, .
\enq

\noindent The two multiple integrals appearing in \eqref{prise en compte de la racine complexe}
 have thus the same gaussian decay. Thus, 

\beq
\lim_{m\tend \infty} \f{\ln \pa{\tau\pa{m}}}{m^2}  = 2 \ln
\paf{\pi}{\zeta} + \mathscr{S}^{(\infty)}
\enq

In particular we recover the results at $\Delta=0$ and $\Delta=\tf{1}{2}$.
As already announced we obtain twice the corresponding bulk constant. Unfortunately the
saddle point method cannot catch the exponential behavior of the
leading asymptotics of $\tau\pa{m}$ and in particular the dependence on the boundary field.

\section{Conclusion}

We have obtained the leading asymptotic behavior of the EFP of the open XXZ
spin-$\tf{1}{2}$ chain for some particular values of the couplings. The $\Delta=\tf{1}{2}$
point seems to have, at least at $h_-=0$, similar symmetry properties to
the bulk model. Thus it should be possible to evaluate to the end the integrals appearing
the generating function of $\sg^z$ correlators at $\Delta=\tf{1}{2}$ in the open chain.
As suggested by the saddle point analysis, the presence of the boundary increases
the long range gaussian decay of $\tau\pa{m}$ by roughly a square factor with respect to the bulk.
It would be interesting to derive some formula for the exponential term $\ex{c m}$ appearing in the  asymptotics of $\ln\pa{\tau\pa{m}}$
in the massless regime. All the more that this terms should exhibit a dependency on the boundary field $h_-$.


\section*{Acknowledgments}

I would like to thank J.-M. Maillet for stimulating discussions.
 K. K. Kozlowski is supported  by the ANR programm GIMP ANR-05-BLAN-0029-01.

\appendix

\section{Appendix}
\label{appendice asymptotique determinant}

In this Appendix we derive the leading asymptotics of a modified Hankel determinant $det{m}{R}$.
The entries of $R$ reads

\beq
R_{ij}= \ex{i\pi \a } \lim_{\eps \rightarrow 0^+}\Int{\Ctr_\eps}{} \dd z \;
\f{z^{j+k-2}  \omega\pa{z}}{\pl_{p=1}^s \pa{z-a_p}}  \; .
\enq

\noindent The integration is carried over a segment and loops encircling some poles
of the integrand:

\beq
\Ctr_\eps= \intff{-1-i\eps}{1-i\eps}
\bigcup\limits_{p\in E}^{} \Gamma_+\pa{a_p} \qquad \e{with} \qquad E \subset \intn{1}{s} \, .
\enq

 In what follows, we assume $a_p \in \mathbb{C}\setminus \intof{-1}{1}$ and $t \not\in E$ whenever
$a_p\in \intof{-\infty}{-1}$. Moreover we restrict to weight functions of modified Jacobi type:
\beq
\omega\pa{z}=\pa{z-1}^{\a} \pa{1+z}^{\beta} g\pa{z}\;\; .
\nonumber 
\enq
\noindent We also assume $g\pa{z}$  holomorphic in an open
neighborhood of \newline$\intff{-1}{1}\bigcup\limits_{p=1}^{s}\paa{a_p}$ and such that
$g\pa{\intff{-1}{1}} \subset \R^+$.

\subsection{Reduction to a  simpler problem}

We first remove the eventual contour integrals from most of the lines of $R$.
For $j<m-s$,  a linear combination of $L_j, \dots,L_{j+s}$ allows to perform the replacement
\beq
z^{j+k} \rightarrow z^{j+k} \pl_{p=1}^{s}\pa{1-\tf{z}{a_p}}
\enq

\noindent in the integrand. For $j \in  \intn{m-s}{m-2}$ one can  only replace
\beq
z^{j+k} \rightarrow z^{j+k} \pl_{p=j+2-m+s}^{s}\hspace{-3mm}\pa{1-\tf{z}{a_p}} \;\; \; .
\enq

\noindent The above transformations lead to
\beq
\det{m}{R}=  \det{m}{\widetilde{R}} \pl_{p=1}^s \pa{-a_p}^{s+1-m-p}  \; \; ,
\enq
\noindent where
\beq
%
%
%
       \widetilde{R}_{jk}= \left\{ \ba{cc}

                \Int{-1}{1} \dd z \, z^{j+k-2}\,  \ov{\omega}\pa{z}
                        &1\leq j \leq m-s\\
                \ex{i\a \pi }\lim_{\eps \rightarrow 0^+}\Int{\Ctr_{\eps}}{} \dd z
                \, \f{z^{j+k-2} \, \omega\pa{z}}{\pl_{p=1}^{j-m+s} \pa{z-a_p}}
                & m-s< j \leq m
    \ea \right.
\enq

\noindent and $\ov{\omega}\pa{x}= \ex{i\pi \a} \omega\pa{x-i0^{+}}$

The second step separates the "pole" part of the determinant from its pure Hankel part.
Let $\Pi_{k}\pa{z}$ be the monoic orthogonal polynomials on $\intff{-1}{1}$ with respect to the weight
$\ov{\omega}\pa{x}$. The reconstruction of these polynomials in the first $m-s$ lines as well as in all the
columns makes the first $m-s$ lines diagonal.
This procedure splits $\det{m}{\widetilde R}$ into
a product of two determinants. Namely,

\beq
\det{m}{\widetilde R}=\det{m-s}{H} \det{s}{K^{(m)}}
\enq

Where

\beqa
H_{jk}&=& \Int{-1}{1} z^{k+j-2} \,\ov{\omega}\pa{z} \dd z \\
K_{jk}^{(m)}&= & \lim_{\eps \tend 0^+} \Int{\Ctr_{\eps}}{}
                \f{z^{j+m-s-1} \Pi_{k+m-s-1}\pa{z}}{\pl_{p=1}^{j}\pa{z-a_p}} \omega\pa{z}\dd z
\eeqa

\subsection{The Hankel Determinant}

The asymptotics of the Hankel determinant can be obtained \textit{via} a procedure established by
Basor and Ehrhardt. They built
a mapping between Hankel and Toeplitz determinants \cite{BasorBijectionToeplitzHankelDeterminant}.
This mapping allowed them to infer the large size behavior  of the Hankel determinant
from the known asymptotics of the Toeplitz determinant. We recall their result in

\begin{prop}
Let T and H be respectively a Toeplitz and a Hankel matrix
\beqa
 T_{kl}&=&\f{1}{2\pi} \Int{-\pi}{\pi} c\pa{\theta} \e{e}^{-i
\pa{k-l} \theta} \dd \theta \\
H_{kl}&=&  \Int{-1}{1} \ov{\omega}\pa{x} x^{k+l-2} \dd x
\eeqa

\noindent such that $\ov{\omega}\in L^1\intff{-1}{1}$ and

\beq
c\pa{\theta} = i \e{sgn}\pa{\theta} \ov{\omega}\pa{\cos \theta} \qquad -\pi<
\theta< \pi \,\, ,
\enq

\noindent then
\beq
\det{m}{H}=2^{-m\pa{m-1}}\pi^{m}\sqrt{\det{2m}{T}}
\enq

\end{prop}

The function $c: \th \mapsto i\e{sgn}\pa{\th} \ov{\omega}\pa{\cos \th}$ is of degenerate Fischer-Hartwig type
\footnote{If $\a$ or $\beta = -\tf{1}{2}$ then $c\pa{\theta}$ is not degenerate. In this case one should drop
the $\pa{-1}^m$ factor in \eqref{Asymptotique Du Toeplitz}. Moreover the result doesn't rely on a conjecture at
these points as asymptotics of Toeplitz determinants with such Fischer-Hartwig symbols are known
 \cite{EhrhardtAsymptoticBehaviorOfFischerHartwigToeplitzGeneralCase}.}
\cite{EhrhardtAsymptoticBehaviorOfFischerHartwigToeplitzGeneralCase} as
it admits two maximal decompositions into the canonical Fischer-Hartwig product:
\beqa
c\pa{\th}&=& 2^{-\a-\beta}h\pa{\cos \th} w_{\pi,\beta,\tf{1}{2}}\pa{\ex{i\th}} w_{0,\a,-\tf{1}{2}}\pa{\ex{i\th}} \nonumber\\
&=& - 2^{-\a-\beta} h\pa{\cos \th} w_{\pi,\beta,-\tf{1}{2}}\pa{\ex{i\th}} w_{0,\a,\tf{1}{2}}\pa{\ex{i\th}} \nonumber
\eeqa

\noindent The function $w_{\th_r,\tau,\sg}\pa{\ex{i\th}}$ reads
\beq
w_{\th_r,\tau,\sg}\pa{\ex{i\th}} = \pa{2-2\cos\pa{\th-\th_r}}^{\tau} \ex{i\sg\pa{\th-\pi-\th_r}} \quad \th_r<\th< \th_r+2\pi
\enq

 The assumptions made on $g\pa{z}$ guarantee the existence of a Wiener-Hopf decomposition:
\beq
g\pa{\cos\theta}= C\pac{g} g_{+}\pa{\ex{i\th}} g_{+}\pa{\ex{-i\th}} \;\; ,
\enq

\noindent where $C\pac{g}$ and $g_+\pa{t}$ satisfy the constraints:
\beq
                    C\pac{g}\in \R   \hspace{2cm}
        g_+\pa{t}= \e{exp}\paa{\sul{n=1}{+\infty} t^n \pac{g}_n} \; ; \;\; \pac{g}_{n}\in \R
\enq

The leading order asymptotics of $H$ thus follow from the generalized Fischer-Hartwig
conjecture raised by Basor and Tracy \cite{BasorTracyGeneralizedFischer-HartwigConjecture}:
\beqa
\det{m}{T_{kl}\pa{c}}=\paf{C\pac{g}}{2^{\a+\beta}}^{m} \f{  E\pac{g} f_{\a,\beta} \, m^{\a^2+\beta^2-\f{1}{2}}}
                            { 4^{\a \beta} g_+^{2\beta}\pa{-1}g_+^{2\a}\pa{1}}
    \pa{\f{1+\pa{-1}^m}{\sqrt{2}} + \e{o}\pa{1}} \;\; ,
\label{Asymptotique Du Toeplitz}
\eeqa

\noindent where
\beqa
f_{\a,\beta}= \pl_{t=\a,\beta}\f{G\pa{1+t+\tf{1}{2}}G\pa{1+t-\tf{1}{2}}}
                {G\pa{1+2t}}  \qquad \e{and} \qquad    E\pac{g}=\ex{\sul{1}{+\infty} n \pac{g}^2_{n}} \; .
\eeqa

\noindent Here G is the Barnes G function.\\

 The asymptotic behavior of the Hankel determinant
reads:
\beq
\det{m-s}{H}=2^{-\pa{m-s}^2} \paf{2\pi C\pac{g}}{2^{\a+\beta}}^{m-s}
\f{m^{\tf{\pa{\a^2+\beta^2}}{2}}\, 2^{\f{\pa{\a-\beta}^2}{2}}}
{m^{\f{1}{4}}\, g^{\beta}_{+}\pa{-1} g^{\a}_{+}\pa{1}}
\sqrt{E\pac{h}\, f_{\a\,\beta}} \; \pa{1+\e{o}\pa{1}}
\label{Asymptotique Hankel}
\enq

\subsection{Asymptotics of $\det{s}{K^{(m)}}$}

Computing the poles and taking the $\eps\tend 0^+$ limit yields

\beqa
K_{jk}^{(m)}&=& \Int{-1}{1}
                            \f{z^{j+m-s-1} \Pi_{k+m-s-1}\pa{z}}{\pl_{p=1}^{j}\pa{z-a_p}} \ov{\omega}\pa{z}\dd z  \\
    && \hspace{3cm}+ 2i\pi \ex{i\a \pi} \sul{t=1}{j} \f{a_{t}^{m-s+j-1} \Pi_{k+m-s-1}\pa{a_t}}
    {\pl_{\substack{p=1 \\ \not=t}}^{j} a_{tp}} \omega\pa{a_t} 1_{E}\pa{t} \; . \nonumber
\label{definition de K(m)}
\eeqa

\noindent Here $1_E$ stands for the characteristic function of E.
The asymptotic analysis of \eqref{definition de K(m)} is based on the uniform
asymptotic estimates for the monoic
orthogonal polynomials with respect to the modified Jacobi weight $\ov{\omega}$
\cite{KuilajaarsMVVUniformAsymptoticsForModifiedJacobiOrthogonalPolynomials}. Let
K be compact in $\mathbb{C}\setminus\intff{-1}{1}$ and $\tilde{K}$ compact in $\intoo{0}{\pi}$. Then :
\beqa
 \Pi_n\pa{\cos\th} &=&  \f{D_{\infty}2^{\tf{1}{2}-n} }{\sqrt{\ov{\omega}\pa{\cos \th}\sin\pa{\th}}}
\cos\paa{\pa{n+\f{\a+\beta+1}{2}} \theta
+\Psi\pa{\th}}\eps_n \quad  \label{asymptotique uniforme [-1;1]} \vspace{3mm}\\
\Pi_n\pa{z} &=& \f{D_{\infty}}{D\pa{\ov{\omega},z}}
            \f{\pa{\tf{\chi}{2}}^{n} }{ \pa{1-\chi^{-2}}^{\tf{1}{2}}} \; \eps'_n
\qquad z=\f{\chi+\chi^{-1}}{2} \,\, , \abs{\chi}>1 \,\, .
\label{asymptotiqueuniforme Complementaire[-1;1]}
\eeqa

\noindent One has $\eps_n, \eps'_n =\pa{1+\e{O}\pa{\f{1}{n}}}$ uniformly in $\theta \in \tilde{K}$ and
respectively $z\in K$. Moreover have introduced the functions:
\beqa
\Psi\pa{\theta}&=&  \f{i}{2} \ln \paf{g_+\pa{\ex{-i\th}}}{g_+\pa{\ex{i\th}}}-\f{\pi}{2} \pa{\f{1}{2}+\a}  \\
D\pa{\ov{\omega},z} &=&  \sqrt{\f{G\pac{g}}{2^{\a+\beta}}}  \, g_+\!\pa{\chi^{-1}} \pa{1-\chi^{-1}}^{\a}
                    \pa{1+\chi^{-1}}^{\beta} \\
D_{\infty}&=& \lim_{\Re{z}\tend +\infty} D\pa{\ov{\omega},z}= \sqrt{\f{G\pac{g}}{2^{\a+\beta}}}
\eeqa

Recall that we have decomposed z into  $z=\tf{\pa{\chi+\chi^{-1}}}{2}$, $\abs{\chi}>1$
and Wiener-Hopf factorized $g$. It follows immediately from $\eqref{asymptotiqueuniforme Complementaire[-1;1]}$
that the leading order of the "pole part" in \eqref{definition de K(m)} reads

\beqa
&&\hspace{-2cm}\f{a_{t}^{m-s+j-1} \Pi_{k+m-s-1}\pa{a_t}}
    {\pl_{\substack{p=1 \\ \not=t}}^{j} a_{tp}} \omega\pa{a_t} =
          \f{C\pac{g}}{2^{\a+\beta}} \f{a_t^{m-s+j-1} \pa{\tf{\chi_t}{2}}^{m-s+k-\tf{1}{2}}}
                {\sqrt{\pa{1-\chi^{-2}_t}}}
    \nonumber \\
 &&\hspace{2cm}\times\;
       g_{+}\pa{\chi_t} \,  \f{\pa{\chi_t-1}^{\a}\pa{\chi_t+1}^{\beta}}
                            {\pl_{\substack{p=1\\ \not= t}}^{j} a_{tp}}
            \eps_n'
\eeqa

\noindent

\noindent We have used the identity

\beq
\f{\pa{\chi_t-1}^{\a}\pa{\chi_t+1}^{\beta}}{2^{\a+\beta}}=
 \f{\pa{a_t-1}^{\a}\pa{a_t+1}^{\beta}}{ \pa{1-\chi_t^{-1}}^{\a}\pa{1+\chi_t^{-1}}^{\beta}}
\enq

We now derive the asymptotics of the integral term in \eqref{definition de K(m)}.
This amounts to the study of the asymptotics $n,q \tend +\infty \, , \, n-q+j\geq 1$, of the sequence

\beq
\mc{I}_{q , n} = \Int{-1}{1}\f{z^q \; \Pi_{n}\pa{z}}{\pl_{p=1}^{j}\pa{z-a_p}} \ov{\omega}\pa{z}\dd z
\enq

Applying the formula for the asymptotics of the orthogonal polynomials \eqref{asymptotique uniforme [-1;1]}
we get:

\beqa
\mc{I}_{q n} & = & \f{D_{\infty}}{2^{n+\f{1}{2}}}\Int{0}{\pi} \dd \th\;
                       \f{\cos^q\th  \, \sqrt{g\pa{\cos \th}}}{\prod_{p=1}^{j}\pa{\cos\th-a_p}}
 \, \pa{1+\cos\th}^{\f{\beta}{2}+\f{1}{4}}\pa{1-\cos\th}^{\f{\a}{2}+\f{1}{4}}\nonumber \\
&& \hspace{1.5cm} \times
\pa{\ex{i\pa{n+\f{1+\a+\beta}{2}}\th + i\Psi\pa{\th}} + \ex{-i\pa{n+\f{1+\a+\beta}{2}}\th  -i\Psi\pa{\th}}}
\eps_n
\eeqa

We recast the integral as one over $\intoo{-\pi}{\pi}$ and then,
 using  the  analycity  of the integrand, we shift the contour to
$\th \in \intoo{-\pi+i0^+}{\pi+i0^+}$. Since

\beq
 \pa{\cos\th-1}^{\gamma}=
            \left\{ \ba{cc}
                \pa{1-\cos\th}^{\gamma}\ex{-i\f{\gamma\pi}{2}} & \th \in\intoo{i0^+}{\pi+i0^+}\\
                \pa{1-\cos\th}^{\gamma}\ex{i\f{\gamma\pi}{2}} & \th \in\intoo{-\pi+i0^+}{i0^+}
            \ea\right.
\enq

\noindent we can absorb the factors $\ex{\mp i\f{\pi}{2}\pa{\a +\f{1}{2}}}$. Moreover 
$2^{\gamma}\pa{\cos \th \pm 1}^{\gamma}=\ex{-i\gamma}\pa{1+\pm \ex{i \th}}^{\gamma} \;
 \th \in \intoo{-\pi+i0^+}{\pi+i0^+}$. Finally, we obtain

\beq
\mc{I}_{q,n} =
\f{C\pac{g}}{2^{n+\a+\beta+1}} \Int{-\pi+i0^+}{\pi+i0^+}
\!\! \f{\pa{1+\ex{i\th}}^{\beta+\f{1}{2}} \pa{1-\ex{i\th}}^{\a+\f{1}{2}} \cos^q \th}
{\pl_{p=1}^{j}\pa{\cos\th-a_p}} \;
\ex{i n\th} g_{+}\pa{\ex{i\th}} \eps_n \,  \dd \th
\enq

\noindent The above integral can be interpreted as a contour integral over $\Gamma\pa{0,1-0^+}$.

\beq
\mc{I}_{q,n}  =   \f{C\pac{g}}{2^{n+\a+\beta+1}} \oint\limits_{\Gamma\pa{0,1-0^+}}^{}
\hspace{-2mm} \f{\dd \zeta}{i\zeta}   \f{\pa{1+\zeta}^{\beta+\tf{1}{2}} \pa{1-\zeta}^{\a+\tf{1}{2}}}
{\pl_{p=1}^{j}\tf{\pa{\zeta+\zeta^{-1} - 2 a_p}}{2}}  \; g_+\pa{\zeta}
\zeta^{n} \paf{\zeta^{-1}+\zeta}{2}^{q} \; \eps_n
\nonumber
\enq

\noindent The integrand has no pole at $\zeta=0$; indeed it behaves as
$\zeta^{j-1+n-p}$, $\zeta \tend 0$, and we have imposed $n+j-p-1 \geq 0$. The only poles inside of the contour
are in the points $\zeta=\chi^{-1}_t\,,\,  t\in\intn{1}{j}$
where $2 a_t= \chi_t+\chi_t^{-1} \,$ and $\abs{\chi_t}>1$. So
\beq
\mc{I}_{q,n} =  \f{2\pi C\pac{g}}{2^{n+\a+\beta}} \sul{t=1}{j}\,  a_t^q\,
    g_+\pa{\chi_t^{-1}}
    \f{\pa{1-\chi_t^{-1}}^{\a+\f{1}{2}}\pa{1+\chi_t^{-1}}^{\beta+\f{1}{2}}}
    {\chi^n_t\pa{\chi_t^{-1}-\chi_t} \pl_{\substack{p=1\\ \not=t}}^{j}a_{tp}} \eps_n \;\; .
\enq

\noindent Thus $K_{jk}^{\pa{m}}$ has the asymptotic behavior

\beqa
K_{jk}^{(m)}&=& \f{2\pi C\pac{g}}{2^{m-s+k+\f{1}{2}}} \sul{t=1}{j}
\left\{ i \ex{i\pi\a }g_+\pa{\chi_t} \chi_t^{m-s+k-1}
\pa{\chi_t-1}^{\a} \pa{1+\chi_t}^{\beta} \, 1_E\pa{t} \hspace{2cm}\right. \nonumber \\
&&\left.       \hspace{1.5cm}-   g_+\pa{\chi^{-1}_t} \chi_t^{s-m-k}
{\pa{1-\chi^{-1}_t}^{\a} \pa{1+\chi^{-1}_t}^{\beta}} \right\}
 \f{a_t^{m-s+j-1} \pa{1+\e{o}\pa{1}}}{\sqrt{1-\chi_t^{-2}}\pl_{\substack{p=1\\ \not=t}}^{j} a_{tp}}
\nonumber
\eeqa

One can simplify the asymptotics of $\det{s}{K^{(m)}}$ by performing the
linear combination of lines:

\beq
L_j \leftarrow L_j - \sul{t=1}{j-1} \f{a_{t}^{j-1+t}}{\pl_{p=t+1}^{j} a_{tp}} \; L_t \qquad j=2 \dots s
\enq

\noindent Then

\beqa
\det{s}{K^{(m)}}&=& \f{2^{\f{s\pa{s+1}}{2}}}{2^{ms}} \paf{2\pi C\pac{g}}{2^{\a+\beta}}^{s}
 \f{1}{\pl_{k<j}a_{jk}} \pl_{t=1}^{s} \f{ a^{m-s+t-1}_t}{\sqrt{1-\chi_t^{-2}}}  \; \pa{1+\e{o}\pa{1}} \hspace{3cm}\\
&& \hspace{-1cm}\times \e{det}_{s} \left[ i \ex{i\pi\a } g_+\pa{\chi_t} \chi_t^{m-s+k-1}
\pa{\chi_t-1}^{\a} \pa{1+\chi_t}^{\beta}
 \, 1_E\pa{t}   \right. \nonumber \\
&&\hspace{2cm} \left. -g_+\pa{\chi^{-1}_t} \chi_t^{s-m-k}
\pa{1-\chi^{-1}_t}^{\a} \pa{1+\chi^{-1}_t}^{\beta}
\right]
\nonumber
\eeqa

\noindent Note that the terms coming from the integration over $\intoo{-1}{1}$
are exponentially sub-leading with respect to the ones coming from the residues.
Nevertheless we keep them in the final formula as we could have $t \not \in E$. For instance,
in the limit $E=\emptyset$ we recover the expected formula \eqref{Asymptotique Hankel} for the asymptotics 
of Hankel determinants with a weight $\tf{\ov{\omega}}{\pl_{p=1}^{j} \pa{z-a_p}}$.

Summing up all the results, we get the formula for the leading asymptotics of $\det{m}{R}$

\beqa
\det{m}{R}&=&2^{-m^2} \paf{2^{s+1}\pi C\pac{g} }{\pa{-1}^{s}2^{\a+\beta}}^{m}
\f{m^{\f{\a^2+\beta^2-\tf{1}{2}}{2}} 2^{\f{\pa{\a-\beta}^2}{2}}}
{g_+^{\beta}\pa{-1}g_+^{\a}\pa{1} \pl_{p>k}^{s}2 a_{kp}}  \;
\sqrt{\f{f_{\a,\beta}E\pac{g}}{\pl_{t=1}^{s} \pa{1-\chi_t^{-2}}} }
\nonumber \\
&&\hspace{-5mm} \e{det}_{s} \left[ g_+\pa{\chi^{-1}_t} \chi_t^{s-m-k}
\pa{1-\chi^{-1}_t}^{\a} \pa{1+\chi^{-1}_t}^{\beta} \right. \label{Asymptotique de Det R} \\
&& \left. -i \ex{i\pi\a } g_+\pa{\chi_t} \chi_t^{m-s+k-1} \pa{\chi_t-1}^{\a} \pa{1+\chi_t}^{\beta}
 \, 1_E\pa{t}  \right]  \pa{1+\e{o}\pa{1}}
\nonumber
\eeqa


\begin{thebibliography}{99}


\bibitem{BetheSolutionToXXX}
H.~Bethe,
"On the theory of metals: Eigenvalues and Eigenfunctions of a linear chain of atoms.",
{\it Zeitschrift f$\ddot{u}$r Physik} {\bf 71}, 205-226 (1931)


\bibitem{Yang-YangXXZproofofBetheHypothesis}
C. N.~Yang and C. P.~Yang,
"One dimensional chain of Anisotropic Spin-Spin interactions: I Proof of Bethe's hypothesis.",
{\it Phys. Rev.} {\bf 150}, 321-327 (1966).


\bibitem{JimKKKM92ElementaryBlocksXXZperiodicDelta>1}
 M.~Jimbo, K.~Miki,  T.~Miwa and A.~Nakayashiki,
"Correlation functions of the XXZ model for $\Delta<-1$.",
{ \it Phys. Lett A} {\bf 168}, 256-263 (1992)


\bibitem{KMTElementaryBlocksPeriodicXXZ}
N.~Kitanine, J.-M.~Maillet and V.~Terras,
"Correlation functions of the XXZ Heisenberg spin-1/2 chain in a magnetic field.",
{\it Nucl. Phys. B} {\bf 567 }, 554-582 (2000)


\bibitem{AlcarazBAopenXXZ}
  F. C.~Alcaraz,  N. M.~Batchelor,  R.J.~Baxter and G. R. W.~Quispel
"Surface exponents of the quantum XXZ,
Ashkin-Teller and Potts models.",
{\it J. Phys. A: Math. Gen.} {\bf 20}, 6397-6409 (1987)


\bibitem{SklyaninABAopenmodels}
E.K. Sklyanin,
"Boundary conditions for integrable quantum systems.",
{\it J. Phys. A: Math. Gen.}  {\bf 21}, 2375-2389 (1988)


\bibitem{JimKKKM95XXZChainWithaBoundary},
 M.~Jimbo,  R.~Kedem,  T.~Kojima, H.~Konno and  T.~Miwa,
"XXZ chain with a boundary.",
{ \it Nucl. Phys.} { \bf B441}, 437-470 (1995)



\bibitem{KozElemntaryblocksopenXXZ},
N.~Kitanine, K.K.~Kozlowski, J.-M.~Maillet, G.~Niccoli, N.A.~Slavnov and V.~Terras,
"Correlation functions of the open XXZ chain I.",
hep-th 07071995



\bibitem{KorepinFirstIntroductionoftheEFP},
V. E.~Korepin, A. G.~Izergin,  F. H. L.~Essler and D. B.~Uglov,
"Correlation Function of the Spin-1/2 XXX Antiferromagnet.",
{\it Phys. Lett A} {\bf 190}, 182-184 (1994)


\bibitem{StronRazuPeriodicEmptinessPiOver3},
A.V. Razumov and  Yu. G.Stroganov,
"Spin chains and combinatorics.",
{\it J. Phys. A: Math. Gen.} {\bf 34}, 3185-3190 (2001)



\bibitem{KMNTPeriodicEmptinessPiOver3},
N.~Kitanine, J.-M.~Maillet, N.A.~Slavnov and  V.~Terras,
"Emptiness formation probability of
the XXZ spin-1/2 Heisenberg chain at Delta=1/2.",
{ \it J. Phys. A: Math. Gen.} { \bf 35}, L385-L391 (2002)




\bibitem{ShiroisihiTakaNishiEFPatDelta=0Periodic}
M.~Shiroishi, M.~Takahashi and Y.~Nishiyama,
"Emptiness formation probability for the one-dimensional isotropic XY model.",
{\it J.Phys.Soc.Jap.} {\bf 70}, 3535-3543 (2001)



\bibitem{WidomSzegoLimitonCircularArcs},
H.~Widom,
"The strong Szegö limit theorem for circular arcs.",
{\it Indiana Univ. Math. J.} {\bf 21}, 277-283 (1971)


\bibitem{KMNTPeriodicEmptinessAsymptAllDelta}
N.~Kitanine,  J.-M.~Maillet, N.A.~Slavnov and  V. Terras,
"Large distance asymptotic behavior
of the emptiness formation probability of the XXZ spin-1/2
Heisenberg chain.",
{\it J. Phys. A: Math. Gen.} {\bf 35}, L753-L758 (2002)





\bibitem{KozResummationsOpenXXZ},
N.~Kitanine, K.K.~Kozlowski, J.-M.~Maillet, G.~Niccoli, N.A.~Slavnov and V.~Terras,
Correlation functions of the open XXZ chain II.",
{\it to appear}.




\bibitem{EhrhardtAsymptoticBehaviorOfFischerHartwigToeplitzGeneralCase}
T.~Ehrhardt, Ph.D thesis,
"Toeplitz determinants with several Fischer-Hartwig singularities.",
{ \it Fakult$\ddot{a}$t f$\ddot{u}$r Mathematik der Technischen Universit$\ddot{a}$t Chemnitz},
{Chemnitz, Germany},
(1997).




\bibitem{KuilajaarsMVVUniformAsymptoticsForModifiedJacobiOrthogonalPolynomials},
A.B.J.~Kuijlaars,  K.T.-R.~McLaughlin, W.~Van Assche and M.~Vanlessen,
"The Riemann-Hilbert approach to strong asymptotics for orthogonal polynomials on [-1,1].",
{\it Advances in Math.} {\bf 188}, 337-398 (2004)


\bibitem{KuperbergSymclassOfA.M.S.}
G.~Kuperberg,
"Symmetry classes of alternating-sign matrices under one roof.",
{\it Ann. of Math. (2)} {\bf 156}, 835-866 (2002)




\bibitem{BasorBijectionToeplitzHankelDeterminant}
E. L.~Basor and T.~Ehrhardt,
"Some identities for dererminants of structured matrices.",
{math-FA/0008075}.




\bibitem{BasorTracyGeneralizedFischer-HartwigConjecture}
E. L.~Basor and C. A.~Tracy,
"The Fischer-Hartwig Conjecture and generalizations. Current problems in statistical mechanics.",
{\it Physica A} {\bf 177}, 167-173 (1991).





\end{thebibliography}

\end{document}